# Development of an Electronic Medical Image Archiving System for Health Care in Nigeria

Olaniyi, Olayemi Mikail
Department of Computer Engineering
Federal University of Technology Minna, Nigeria

Omotosho Adebayo and Robert Jane
Department of Computer Science and Technology
Bells University of Technology, Ota, Nigeria

Oke Alice.O
Department of Computer Science and Engineering
Ladoke Akintola University of Technology, Ogbomoso, Nigeria.
e-mail: alitemi2006 {at} yahoo.com

*Abstract*—**Medical images require immediate access by several physicians in different places within a medical facility and access to a critically injured person's medical image, such as the x-ray, on time can be a key factor in the diagnosis and treatment of the patient. The electronic medical image archive system can help to solve the problem faced in previous physical medium archiving, thus increasing productivity and time which patients are attended to. In this paper, simple but functional electronic medical image archive architecture was proposed and implemented. The system was further evaluated in a hospital setting by medical experts using sample patient image data. Results of the system evaluation shows that electronic medical image archiving systems can actually promote efficiency, quality improvement, provide timely availability of radiologic images, image consultation, and image interpretation in emergent and no emergent clinical care areas among other benefits**

*Keywords- Telemedicine, Image archiving, medical images, CT Scans and Mammography*

## I. INTRODUCTION

Medical imaging is the technique used to create images of the human body (or parts and function thereof) for clinical purposes [6]. Although, there is an exemption to this definition as the imaging of removed organs and tissues can be performed for medical reasons but such processes are not generally referred to as medical imaging, rather are a part of pathology. Previously, physicians have found it impossible to diagnose internal problems of a patient without cutting open the patients however; medical images provide a medium of diagnoses by accessing the human body through the creation of the human image. Medical imaging is a rapidly expanding field, it started with X-rays, Ultrasonography, after which three-dimensional imaging of anatomical structures was made possible with Computed Tomography (CT) and Magnetic Resonance Imaging (MRI) scans. Newer imaging technologies such as the Functional magnetic resonance (FMRI), Position emission tomography (PET), and Single Photon Emission Computed Tomography (SPECT) scanning allows creating images of body function rather than anatomy [2]**.**

Medical imaging, especially X-ray based examinations and ultrasonography, is crucial in every medical setting and at all levels of heath care. In public health and preventive medicine as well as in curative medicine, effective decisions depend on correct diagnosis. Though medical judgment may be sufficient in treatment of many conditions, the use of diagnostic imaging services is essential in confirming, correctly assessing and documenting course of the disease as well as in assessing response to treatment. It plays an important role to the improvement of public health because it comprises of different imaging modalities and processes to image human body for diagnostic and treatment purposes.

Electronic medical image archiving approach produce significantly faster results, patients experience quicker appointments and less anxious waiting for diagnosis. Patients are also likely to appreciate that digital X-rays can achieve high picture quality with a lower dosage of radiation, which means less exposure and less risk. In addition, digital images can easily be transferred from one doctor's office to another via email, on disc or by other electronic means. This saves the patient the difficulty of trying to get physical records transferred, for example from a primary doctor to a specialist or hospital [13]**.** In addition, by storing images on paper and film, there is of course the risk of environmental deterioration and damage, as well as the risk of complete destruction and loss from natural disasters like earthquakes, floods, and fires. Through the digitization of images, there is a reduction storage space and high speed in image film search, hence a reduction in costs for medical institutions. Digital equipment offers medical professionals a sharper image to use in diagnostic procedures; images can also be digitally manipulated using contrast and brightness alterations in ways that are impossible with traditional method. Images are archived into a digital media which reduces storage space. Images can be backed up which increases safety.

The previous systems used to archive images at physical medium such as films have drawbacks as the communications and the analysis of these images were impossible by wider auditorium. The growth in medical archives and the need for long term data retention are problems faced with these physical medium. In the process of digitalization of the old medical image archives, the introduction of ICT has made it possible to provide efficient access to the longstanding databases. Hospitals and other health institutions can use this new technology without spending extra money for buying additional new equipment for analysis and further research. The introduction of ICT to medical image archiving will bring many benefits to hospitals and other health institutions, increasing the efficiency in medical image processing, as well as efficient archiving and retrieving medical databases [3]**.**
.

II. RELATED WORK

In the traditional approach to medical image archiving, images of patients are taken for diagnosis and the films are archived using a physical storage system. When the images are needed by the physician for diagnosis of a patient or reference on previous disease, the film is restored from the archive manually. Conventionally, technicians take an image, radiologists examine it and present their conclusion, and then the image is archived. If a patient was undergoing long term treatment, image could be retrieved weeks or months later to be used to monitor patient's progress. This system is a costly value considering the educational and administrative use of medical images [8]**.** With the adoption of ICT the overhead cost of storing and retrieving images could be reduced. Authors in [6] and [11] emphasized that the growth of computers and image technology, medical imaging has greatly influenced the medical field, and that the diagnosis of a health problem is highly dependent on the quality and the credibility of the image analysis.

However as every new technology always face some setbacks; [5] discussed the challenges that have hindered medical images digitalization for years. The authors stressed the importance of imaging information in health care they concluded that the healthcare process can only become more effective and efficient when the appropriate information is in the right place at the right time, something that conventional methods, using photos that need to be physically moved, can scarcely satisfy. Also, authors in [1] discussed the migration of medical images in developed countries into digital format. The availability of clinical medical images is larger than ever, but the exploitation of its knowledge is poor. Hospitals, clinics and primary care centers store the information locally with a poor level of capability of usage of this information.

The ultimate goal of digital imaging is to become a filmless environment as a result of related improvement in costs and patient care. The results of some pilot projects have shown convincing benefits of digitizing medical images for example, in [7] evidence of the impact of Picture Archiving and Communication Systems (PACS) on clinicians' work practices in the intensive care unit was assessed. Eleven studies from the USA and UK were included; the authors found that the potential for PACS to impact positively on clinician work practices in the Intensive Care Unit (ICU) and improve patient care is great.

Similarly in [8], a research carried out at university of California Davis health system (UCDHS*)* physicians earlier reported they used up to one to three hours looking for films in the night and weekends. But after implementation, the total search time reduced to one hour for all patients. Also, 65% of physicians reported checking out images from libraries making it unavailable for other physicians, some took it for conference, some to show their attendance, some to show their patients, some hide images for personal use. Images were out of radiology departments for an average more than 20 hours. With implementation of a digital image archiving system, these behaviours ceased. After implementation, images were readily available and increased satisfaction was reported. Average image search time decreased, from 16 minutes to 2 minutes, saving 21.5 physician years*,* worth $1,034,140 annually. System implementation saved 21% physician years and 2 million annually.

The design of an acceptable image archiving system therefore becomes paramount in order to meet the expected needs of the intended users. Authors in [4] and [10**]** cited how an effective, low cost, long term archive is essential to the successful implementation and acceptability of a PACS system and any filmless plan. Performance and capacity requirements must be carefully understood and cost/performance tradeoffs must be made in light of these requirements. According to them, many hospitals have recently implemented high capacity archives and are now enjoying the advantages of access to all of their images, without human intervention. The article also describes that for multi-terabyte medical image archives, automated near line archive systems are the only economical way of storing massive amounts of data required for this application. In the same vein, authors in [3] proposed the development of systems for medical images, the type of information that is important to record and archive into the database. Also, the proposed system will be able to offer access to medical database with information about patients, independently of their location, raising health care on a higher level especially in rural areas.

Finally, according to [9], a large percentage of health care professionals and community users in developing countries are not fully ICT aware and most of the approaches being used in the health practice are still at a relatively new stage of implementation, with insufficient studies to establish their relevance, applicability or cost effectiveness and this makes it difficult for governments of developing countries to determine their investment priorities. From the reviewed work process, it was concluded that the system to be developed will be beneficial to most hospitals in Nigeria due to the fact that images will be digitized and there should be a pro rata improvement in the rate which patients are attended to. Also as established from literature, capturing of images into systems will reduce film search and improve rate of access on a patient's information.

### III. SYSTEM DESIGN

The purpose of system design is to create a technical solution that satisfies the functional requirements for the system. The functional specification produced during system requirements analysis is transformed into a physical architecture through system modeling and database design

A. Infrastructural Modeling and Architecting

(i) Overall System Architecture

The architecture as shown in figure 1 describes how the system operates. The patient's image is uploaded into the system by the radiographer. When a patient visits the hospital, the physician logs into the system and pulls patient images/history. The patient server manages patient images uploaded by the radiographer; these images are X-ray, CTscan and Mammography. The cloud hosts the server, and the physician gets access to images by logging into the system which is connected to the web. The physician attends to patient and treatment is given. If there is need to access patient's image in the future, the image can be retrieved in minutes and patient gets treated on time. The system consists of the login authentication function that verifies the users who log onto the system, a centralized database server where patient data is stored. The Application software allows the users of the system to perform the basic operations, i.e. administrator to add patient, delete patient, add users, delete users, assign roles to users,(only administrator can assign roles). The physician accesses patient history, radiographer uploads images, while other users have limited access as assigned to them by the administrator.

Figure 1: Architecture of Proposed System

(ii) Analysis of Model

The structure of the proposed system was analyzed using the Use-Case diagram, Class diagrams and the behavioural/sequence diagram using object oriented paradigm. The use case scenario of the proposed system is shown in Figure 2.The administrator controls the activities of the system and has major functions to the system. The physicians have some limited functions for example, they are not allowed to add or delete patients.

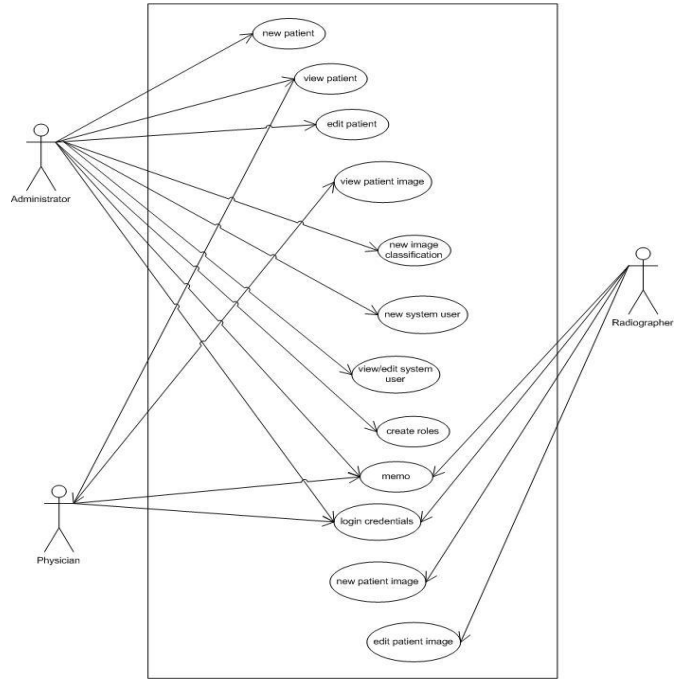

Figure 2.Use case diagram for patient registration

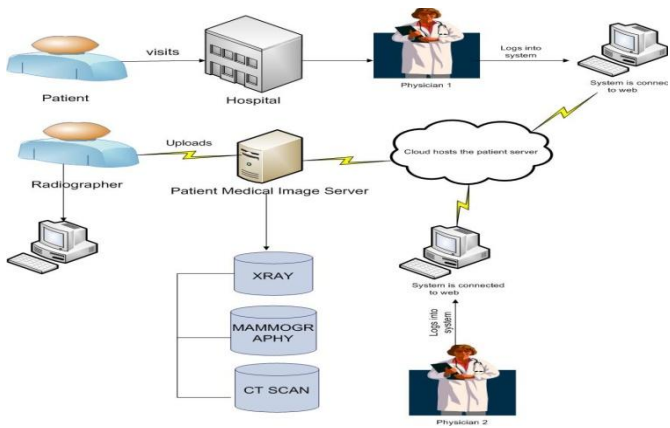

In figure 3 the sequence diagram shows how the system objects communicate with each other in terms of a sequence of messages and also indicates the life spans of objects relative to those messages. The diagrams below show the various processes needed in the system in the form of sequences ranging from user registration, report generation etc. The diagram above shows the way new patients are registered into the database. The administrator is the only user granted access to this part of the system.

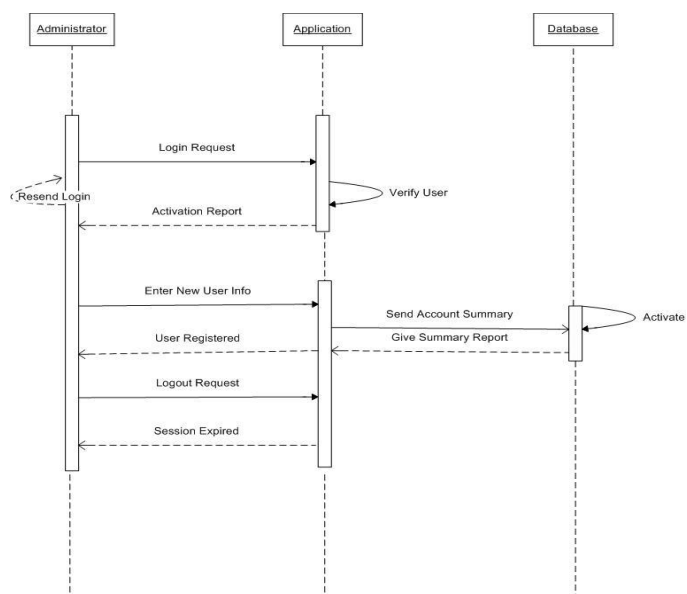

Figure 3 Sequence diagram for patient registration

Figure 4 shows the system's sequence diagram. The diagram depicts the step for identifying patient within the system. The administrator/physician can type in any search letter, word, or category and the system provides the different patient search results that may be important to the finder

### B. Database Design

The electronic image archiving system has eight tables as described below:
**SCAN_CATEGORIES(Scan_Category_Id,**Category_Name ,Category_Description)
**SCANS (Scan_id,** Patient_Id, Scan_Category_ID, Radiographer, Scan_Image, Scan_Timestamp, Expiry, Scan_Details, Comments)
**PATIENT (Patient_Id,** First_Name, Last_Name, Address, Phone, Email, Sex, Card_Number, Photo)
**AUDIT_TRAIL (Log_Id,** User_Id, Event_Description, Event_Timestamp)
**USER_ACCOUNTS (User_Id,** User_Password, Title, First_Name, Last_Name, Sex, Phone, Email, Address, Photo, User_Profession, Account_Status)
**SYSTEM_PRIVILEGES (Privilege_Id,** Privilege_Description, Status)
**ROLE (Role_Id,** Role_Name, Status)
**ROLE_PRIVILEGES (s/n, Role_Id,** Privilege_Id)

### IV RESULTS AND DISCUSSION

Figures 5, 6, 7, 8, 9 and 10 show the impact the developed system using ICT technologies could have on medical images storage and retrieval.

The system login page shown in figure 5 welcomes the users to the system. It gives an overview of what the system should look like. The login page allows the user to fill in the password and username if the user is an existing user, if not the user meets the administrator for a pass. In the case of a mismatch in username and password, the system will trigger an error " login error, check your password and username" in this case, the user should recheck the password and try logging in again.

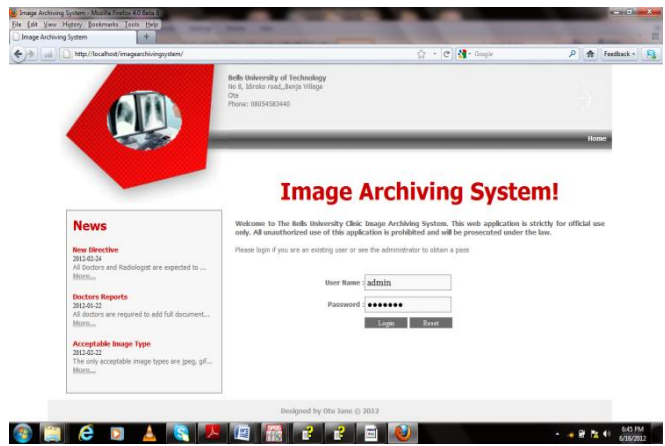

Figure 5. Login Page

Figure 6 allows a new patients image to be uploaded for an already registered patient. The format in updating a new image is as follows: patient card number, scan category(CT, Xray and Mammography), the radiographer, patient scan (uploaded from the system), scan date, scan details, radiographers findings.

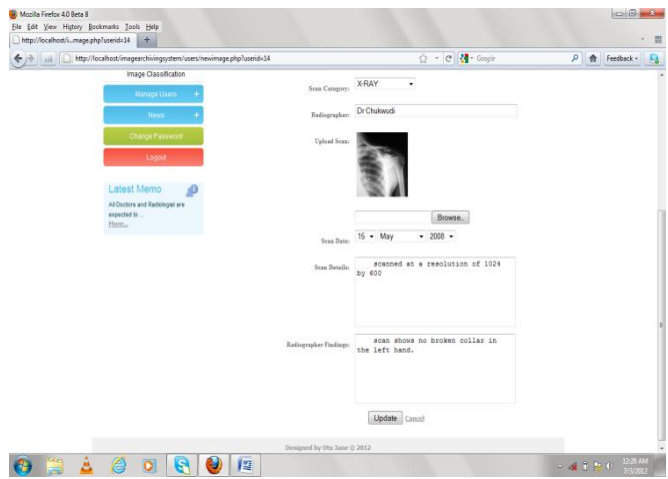

Figure 6. New Patient with X-ray Image page.

The image search page allows medical images of patients to be accessed. When the "view images" link is clicked, a search image page comes up which shows how the images could be searched within the system through different criteria. Search categories include by: scan, radiographer, patients name, patients card number, patient last name, patient first name, patient email and patient phone number. Figure 7 shows the process of searching for a patient medical image information using "radiologist" name (e.g. Dr. Akpan) as the search criteria. Searched images can then be viewed and used for diagnosis purposes.

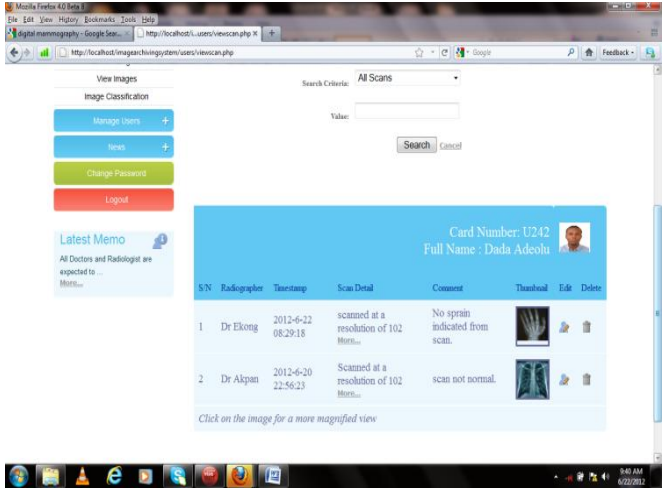

Figure 7. Patient image View.

Figure 8 shows how images are classified either into mammography, ct-scans or x-rays. The images that have already been classified are also listed on this page for viewing and editing. To classify a new image, type the classification name, and the description of the image classification. To edit each classified image, click on edit, update the necessary information and click on update to save.

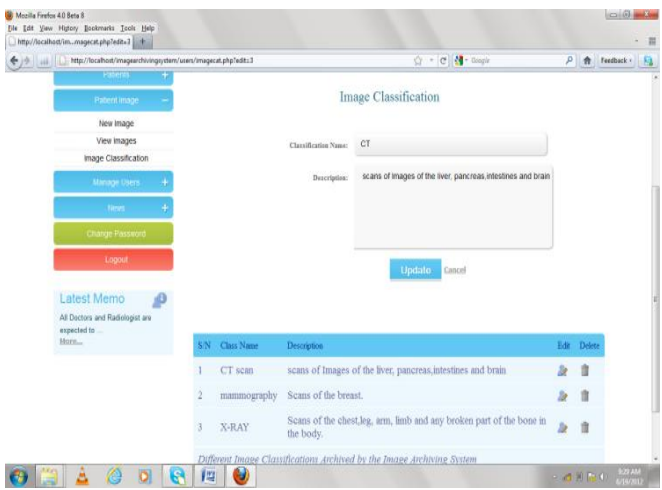

Figure 8. Image classification page.

This screenshot shown in Figure 9 allows the users of the system to be assigned roles and priviledges which means all users cannot have access to some information in the system. Therefore, only users that have been assigned a priviledge can access information based on the functions assigned to those privelegdes. Role priviledge include patients, patient images, manage users, news. Figure 9 shows the role assigned to nurses, the role name is filled as "Doctors" the role description is filled. The role privelege patients, patient's images and news was ticked which means the doctor can only have access to all these but cannot manage users.

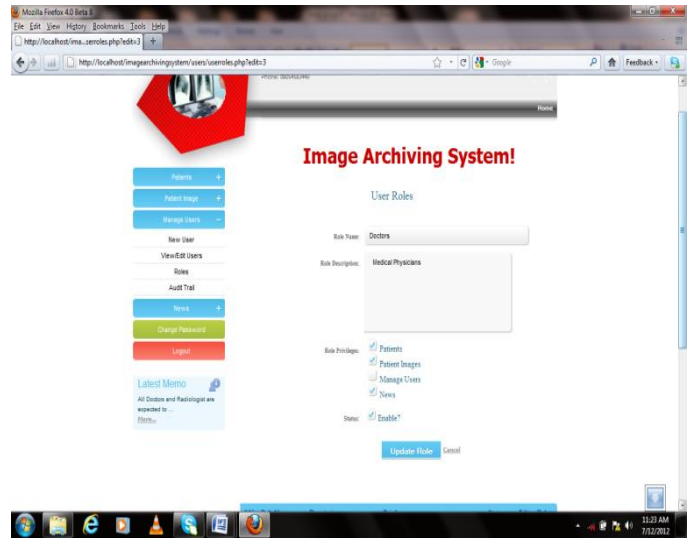

Figure 9. User role page.

The system interface shown in Figure 10 allows viewing all the users on the system and to edit each user information if necessary. To edit, click on edit, change necessary information and save. Only the administrator has the function of changing user information in the system.

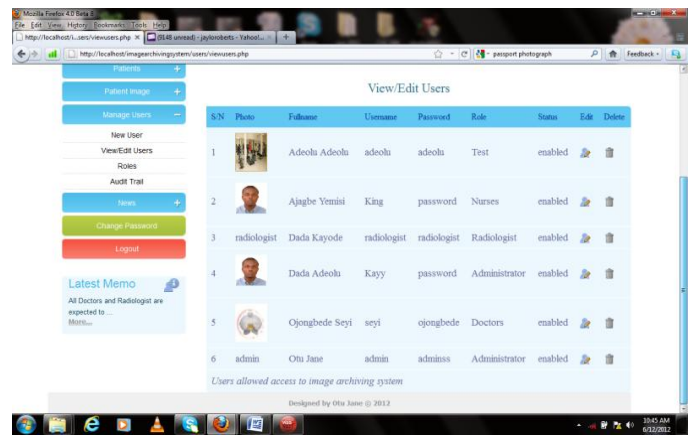

Figure 10. View/edit user page.

## V  SYSTEM PERFORMANCE EVALUATION

The performance evaluation of the developed system was carried out by administering questionnaires to the intended users (physicians) of this system in Nigerian hospitals,precisely Medicare Hospital, Ota and Coscharis Clinic , Ota Ogun State Nigeria. The system was rated against the features of an electronic medical image archive system. The features are represented on the horizontal bar and Ratings on the vertical bar using 5-point linkert items as: 5-Strongly Agree, 4-Agree, 3-Undecided, 2-Disagree, 1-Strongly Disagree. The bar chart in Figure 11 shows the pictoral representation of the evaluation result.

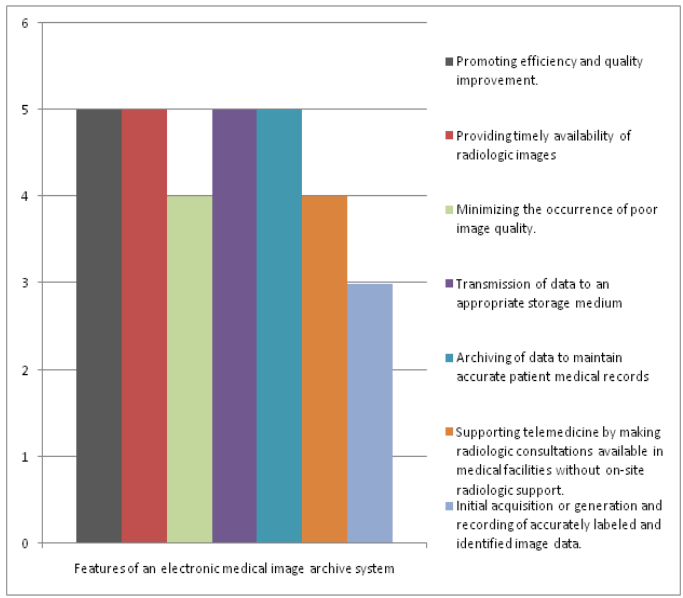

Figure 11: System Performance Evaluation

### VI  CONCLUSION AND RECOMMENDATION FOR FUTURE WORK

The results of the developed system design and implementation shows that digitization of medical images can significantly reduce unauthorized access of images, reduce the manual labor involved in retrieving physical images, and also increase the operational speed of medical health delivery in developing countries like Nigeria. The electronic medical image archive system only covers a limited number of operations such as registration of patients, uploading of images already captured; retrieving of images etc. It is recommended that future research include security features to prevent image misuse.